\documentclass{aa}
\usepackage[dvipsnames]{xcolor}
\usepackage[breaklinks, colorlinks, citecolor=Cerulean]{hyperref}
\usepackage{color}

\usepackage{graphicx}
\usepackage[flushleft]{threeparttable}
\usepackage{txfonts}

\newcommand{\OIII}{[O~{\sc iii}]\ }
\newcommand{\NII}{[N~{\sc ii}]\ }
\newcommand{\HII}{H~{\sc ii}\ }

\newcommand{\HI}{H~{\sc i}\ }
\newcommand{\Ha}{H$\alpha$\ }
\newcommand{\kms}{\,\mbox{km}\,\mbox{s}$^{-1}$}

\newcommand{\SIIHa}{[S~{\sc ii}]/H$\alpha$}

\newcommand{\OIIIHb}{[O~{\sc iii}]/H$\beta$}
\newcommand{\OIIHb}{[O~{\sc ii}]/H$\beta$}
\newcommand{\sunn}{$_{\odot}$}
\newcommand{\atoms}{atoms~cm$^{-2}$}

\def\rev{\textcolor{black}}

\begin{document} 

\title{Chemodynamic evidence of pristine gas accretion in the void galaxy VGS 12}

   \author{Evgeniya Egorova
          \inst{1}
           \and
           Kathryn Kreckel\inst{1}
           \and
           Oleg Egorov\inst{1}
           \and
           Alexei Moiseev\inst{2,3}
           \and
           Miguel A. Aragon-Calvo\inst{4}
           \and
           Rien van de Weygaert\inst{5}
           \and
           Sergey Kotov\inst{2}
           \and
           Jacqueline van Gorkom\inst{6}
          }

   \institute{Astronomisches Rechen-Institut, Zentrum f\"{u}r Astronomie der Universit\"{a}t Heidelberg, M\"{o}nchhofstra\ss e 12-14, D-69120 Heidelberg, Germany\\
              \email{e.egorova@uni-heidelberg.de}
        \and
        Special Astrophysical Observatory, Russian Academy of Sciences, Nizhny Arkhyz 369167, Russia
        \and
        Lomonosov Moscow State University, Sternberg Astronomical Institute, Universitetsky pr. 13, Moscow 119234, Russia
        \and
        Instituto de Astronom{\'i}a, Universidad Nacional Aut{\'o}noma de M{\'e}xico, Apdo. Postal 106, Ensenada 22800, BC, Mexico
        \and
        Kapteyn Astronomical Institute, University of Groningen,PO Box 800, 9747 AD, Groningen, The Netherlands
        \and
        Department of Astronomy, Columbia University, Mail Code 5247, 550 West 120th Street, New York, NY 10027, USA
             }

   \date{Received ; accepted }

  \abstract
   {Accretion of metal-poor gas is expected to be an important channel of gas replenishment in galaxy evolution studies. However, observational evidence of this process is still relatively scarce.}
   {The unusual polar disk galaxy VGS~12 was found in the Void Galaxy Survey. It appears to be isolated and resides in the cosmological wall between two large voids. The suggested formation scenario for this peculiar system is accretion of metal-poor gas from the void interior. To confirm or refute the accretion scenario, information on the chemical properties of VGS~12 is crucial and provides novel insights into gas accretion mechanisms when combined with kinematic constraints on the polar material.}
   {We present for the first time the data on the gas-phase chemical abundance of VGS~12 obtained with the Russian 6m telescope BTA. We complement our analysis with HI data obtained with VLA and the data on the kinematics of the ionized gas.} 
   {VGS~12 appears to be a strong outlier from the ``metallicity -- luminosity'' relation, with gas oxygen abundance $\sim$0.7~dex lower than expected for its luminosity. The nitrogen abundance, on the other hand, is higher than what is typically observed in galaxies with similar metallicity, but is consistent with the metallicity expected given its luminosity. Such behavior is what is expected in the case of metal-poor gas accretion.  
   The H~\textsc{i} reveals clear morphological and kinematical asymmetry between the northern and southern parts of the disk, which are likely related to its unsettled state due to the recent accretion event. The kinematics of the ionized gas seen in \Ha  reveal prolate rotation and follow closely the rotation of the H~\textsc{i} disk, so we suggest this is accreted H~\textsc{i} gas ionized by the stars in the central region of the galaxy.}
   {Together, our findings provide strong, multiwavelength evidence of ongoing cold gas accretion in a galaxy caught in the act of growing from the cosmic web. This is one of the very few individual galaxies where a convincing case can be made for such a process, and demonstrates the potential for cold accretion to contribute to galaxy growth even in the low-redshift universe.}


   \maketitle

\section{Introduction}

Galaxy evolution is governed by star formation, which leads to the consumption of the galaxy’s gas reservoir, but with depletion times less than the age of the universe it would stop quickly \citep[e.g.,][]{Bigiel08, Leroy2008, Bigiel2011, Leroy2013, Lilly2013}. One of the key open questions in modern studies of galaxy evolution is: how do galaxies get their gas, which is necessary for the continued fueling of their star formation activity? In different papers, several mechanisms of galaxy growth and gas replenishment are proposed, such as mergers \citep[e.g.,][]{LHuillier2012,DiMatteo2008}, galactic fountains \citep[e.g.,][]{Fraternali2006,Fraternali2008,Marinacci2010}, and cold gas accretion from filaments \citep{Semelin2005,Keres2005,Dekel2006,Dekel2009}. A number of simulations have shown that smooth accretion from filaments dominates over mergers \citep{LHuillier2012,Wang2011,vandeVoort2011}. However, direct observational evidence of filamentary gas accretion has remained limited. 

According to simulations, cold accretion of pristine gas from the cosmic web was the most important channel for gas replenishment in the early Universe, and it is predicted that it may still proceed nowadays in voids \citep[e.g.,][]{Aragon2013, Aragon-Calvo2019}. The existence of underlying large-scale filaments and walls in voids and indirect indications of gas accretion from the intergalactic medium were indeed suggested in several observational studies, and predicted by theoretical studies and simulation of the hierarchical buildup of the cosmic web \citep{Bond1996,Sheth04}. Previously, two systems with the linear alignment of their components, VGS~31 \citep{VGS2013} and VGS~38, were found in the course of the Void Galaxy Survey \citep{VGS_pilot_Kreckel2011, VGS_full_Kreckel2012}. \cite{rieder13} simulated VGS~31 and concluded that the galaxies in the system formed in the same protofilament. Similar systems with linearly aligned components that may reflect the underlying filamentary structures were found in other observational studies, e.g., J0723+36 \citep{CP2013} and UGC3672 \citep{CPE2017}. The SARAO MeerKAT Galactic Plane Survey \citep{Goedhart2024} has also revealed two groups in the Local Void that show signs of filamentary substructure \citep{Kurapati2024}. 

Combining information on morphology (such as the presence of extended gaseous reservoirs or filamentary structures), gas kinematics (i.e., signatures of noncircular gas motions or misaligned disks) and chemical abundances (indications of the ISM diluted by metal-poor gas) would provide compelling observational support for the cold gas accretion scenario \citep{SA2014}. 
\cite{VGS_full_Kreckel2012} found that many galaxies in their Void Galaxy Survey have strongly disturbed H~\textsc{i} gas morphologies and kinematics. The authors interpret this as signs of ongoing interactions and gas accretion. The disturbed H~\textsc{i} disks were also revealed in the sample of extremely poor galaxies in voids in \cite{Kurapati2024_XMP}. \citet{Egorova2019, Egorova2021} identified localized noncircular motions of the ionized gas that coincide with the peculiarities of chemical abundance in several regions of NGC~428 and Ark~18 void galaxies, which provide indirect support for recent gas accretion. However, none of the existing studies of void galaxies have provided a smoking gun evidence of ongoing metal-poor gas accretion so far. 

The outstanding polar disk galaxy VGS~12 was found by \cite{Stanonik2009} in the Void Galaxy Survey \citep{VGS_pilot_Kreckel2011, VGS_full_Kreckel2012}.  This galaxy is isolated and situated in a thin cosmological wall between two voids. H~\textsc{i} disk of VGS~12 is perpendicular to the central stellar disk, has no optical counterpart, and is almost perpendicular to the wall where the galaxy resides. It is generally considered that polar disk or polar ring galaxies (PRG) are formed as the result of the interaction between a host galaxy and its companion (either mergers or the gas accretion from the companion), or due to cold accretion from gaseous filaments. \cite{Stanonik2009} discuss the latter scenario, i.e., the cold gas accretion from the adjacent void, as the most likely formation mechanism for the H~\textsc{i} polar disk in VGS~12. There is no available information on the chemical properties of VGS~12, although it would be particularly crucial to verify the pristine gas accretion scenario. 

\begin{figure*}\centering	\includegraphics[width=1.6\columnwidth]{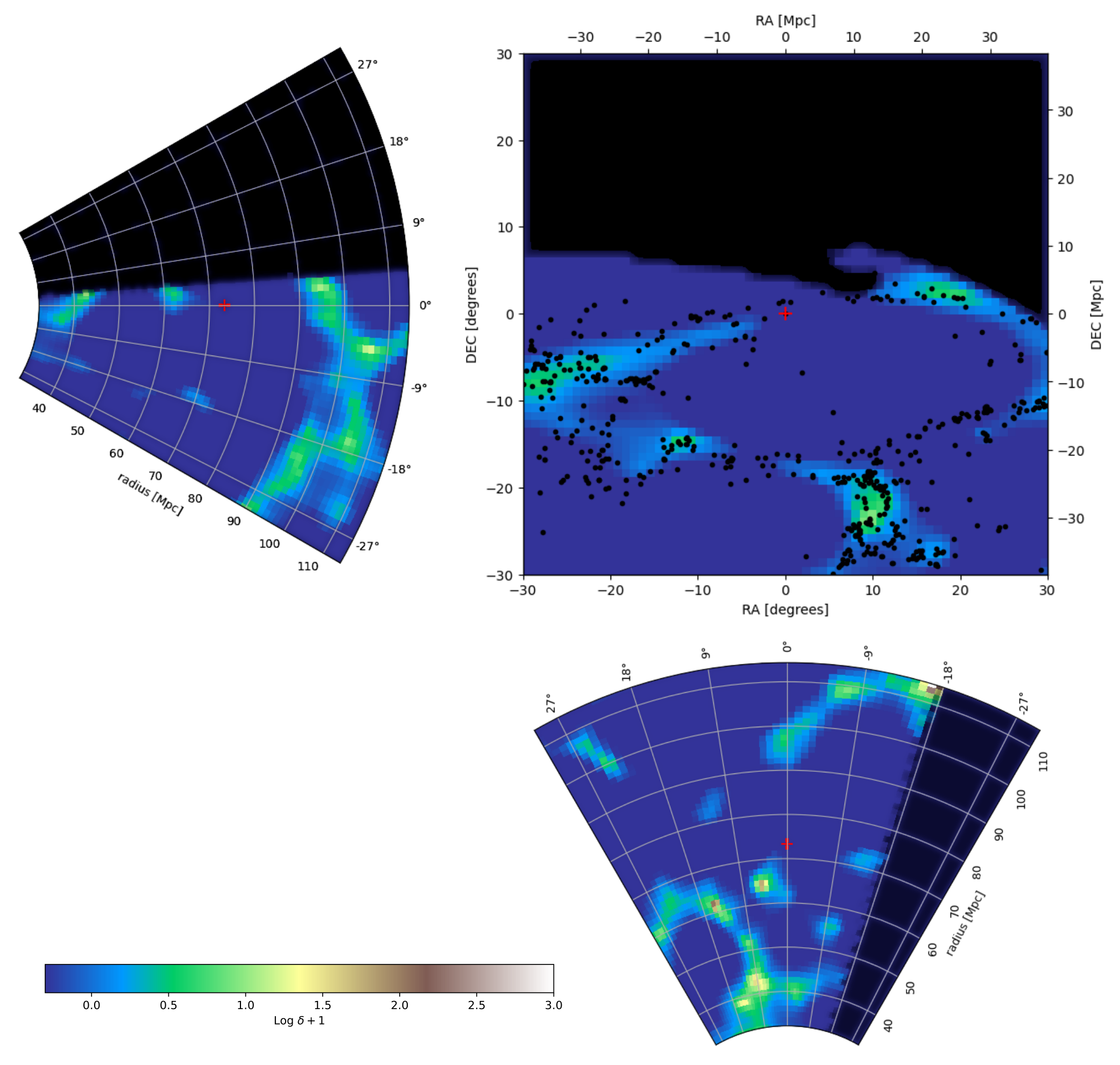}
  \caption{Cosmic structures in the vicinity of the galaxy VGS~12. Three orthogonal cuts across the density field are shown (clockwise): $z$-Dec,  RA-Dec, RA-$z$ (RA and Dec centered on VGS~12). The red cross marks the position of VGS~12. In the RA-Dec slice we show galaxies in a $\pm10$~Mpc slice in $z$ as black dots. The density field $\log (1 + \delta)$ is shown in the background. The regions colored in black are located outside the SDSS mask.
  }
  \label{fig:environment}
\end{figure*}

We show the cosmic structures in the vicinity of VGS~12 galaxy in Fig.~\ref{fig:environment}. In order to get a better view of the cosmic environment around VGS~12 we used a visualization tool that creates slices of the density field around a given galaxy (Aragon-Calvo in prep.). The density field was computed using the Stochastic Delaunay Tessellation Field Estimator \citep{Aragon21}, which improves the Delaunay Tessellation Field Estimator algorithm \citep{schaapwey2000,weyschaap2009} producing a smooth reconstruction. Three orthogonal cuts ($z$-Dec,  RA-Dec and RA-$z$) are made across the density field centered on the target galaxy. We can see that although VGC12 is near the edge of the SDSS mask, the surrounding LSS is consistent with a wall separating two voids. This is more evident in the RA-Dec slice in which VGS~12 is shown between two elongated and tenuous structures corresponding to the edges of its parent wall. 

In this paper, we present a new analysis for VGS~12 exploiting previously unpublished Karl G. Jansky Very Large Array (VLA) data together with long-slit spectroscopic and Fabry-Perot interferometer data (in the \Ha line) obtained at the  6m telescope BTA of the Special Astrophysical Observatory Russian Academy of Sciences (SAO RAS). We provide the main properties and parameters derived in this paper in Table~\ref{tab:summary}. 
The paper is organized as follows. We provide the overview of the observational data and data reduction in Section~\ref{sec:data}, then outline the main results on the morphology and kinematics of gaseous disk and chemical abundance in Section~\ref{sec:results}, and discuss the results in the context of the evolution of VGS~12 and the processes that take place in voids in Section~\ref{sec:discussion}, followed by the conclusions in Section~\ref{sec:conclusions}.

\begin{table}
	\centering
	\caption{Main properties and derived parameters of VGS~12. 
 }
	\label{tab:summary}
\begin{tabular}{l|c} \hline   \\ [-0.2cm]
Parameter & Value \\
\\[-0.2cm] \hline \\[-0.2cm]
RA	(J2000)$^a$	    & 10:28:19.24   \\
Dec	(J2000)$^a$	    & 62:35:02.60	\\
D$^b$, Mpc          & 76  \\
$M_{HI}^b$ 	& $3 \times 10^9$	\\
$M_*^c$           & $1.05 \times 10^9$  \\
$M_B^d$           & -19.29  \\
$V_\mathrm{hel}^e$, \kms &  5295\\
$i^e$ , deg 		    &  64	\\
$12+\log\mathrm{(O/H)}^e$ &  $7.67\pm0.07$\\
\hline %
\end{tabular}
\\[0.2cm]
\begin{tablenotes}
\item $^a$ From NASA/IPAC Extragalactic Database (NED)
\item $^b$ \cite{Stanonik2009}
\item $^c$ \cite{VGS_pilot_Kreckel2011}
\item $^d$ \cite{VGS_Beygu_2017}
\item $^e$ This work
\end{tablenotes}
\end{table}

\section{Observations and data reduction}
\label{sec:data}

\subsection{VLA data}
\label{sec:data_VLA}

We use archival VLA data (PI: K.Stanonik), obtained in B-configuration, which allows us to achieve significantly higher resolution (up to $\sim$~5$\arcsec$) than the WSRT $\sim$~20$\arcsec$ data presented in \cite{Stanonik2009}.
Total bandwidth 3.125 MHz was divided into 128 channels (velocity resolution of $\sim$5~km~s$^{-1}$). The target was observed for a total of 14 hours. The parameters of the VLA observations are summarized in Table~\ref{tab:VLA_obs}.

Calibration and flagging were performed following standard procedures in the Astronomical Imaging Processing System (AIPS). The data were corrected for the Doppler shift due to Earth’s motion. The target was observed over two sessions (on March 30 and April 04, 2009), which were combined using the task ``DBCON'' before imaging. Continuum emission was removed from the UV data by linearly interpolating over the line-free channels. The data cubes were generated using a CLEAN box placed around the H~\textsc{i} emission. They were produced for various (u,v) ranges, including 0–10, 0–20, and 0–40 k$\lambda$, corresponding to a beam full width at half-maximum (FWHM) of roughly $\sim$~17$\arcsec$, $\sim$~9$\arcsec$ and $\sim$~5$\arcsec$, respectively. Images were created with robust weighting, zeroth and first moment maps were made using Hanning and Gaussian smoothing to 3~km~s$^{-1}$ and 7$\arcsec$, respectively.

\begin{table}
	\centering
	\caption{Parameters of the VLA observations.}
	\label{tab:VLA_obs}
\begin{tabular}{l|c} \hline  \\ [-0.2cm]
 & VGS~12 \\
\\[-0.2cm] \hline \\[-0.2cm]
Configuration			&   B \\
Dates of observations & 2009 Mar 30, Apr 04	\\
Field center R.A.(2000)	& 	10h28m19.0s	\\
Field center Dec.(2000) &  62$^{\circ}$35'2.0" \\ 
Central Velocity (\kms)  &  5330	\\
Time on-source (h) &  14\\
Number of channels & 128 \\
Channel separation (\kms) & ~5.2 \\
Flux Calibrators &  1331+305 \\
Phase Calibrators & 1035+564 \\
\hline 
	\end{tabular}
\end{table}

\subsection{Observations with scanning Fabry--Perot interferometer}
\label{sec:fpi}

\begin{table*}
\centering
\caption{\rev{Observations at 6-m Russian telescope, with long-slit and scanning Fabry-Perot interferometer at SCORPIO-2 multimode focal reducer}}
\label{tab:Obs}
\begin{tabular}{lccccccc} 
\hline
{Data set} & {Date of obs.} & {$T_{exp}$, s}& {$FOV$} & {Scale, arcsec~px$^{-1}$} & {Seeing, $\arcsec$} & {Spectral range, \AA} & {$\delta\lambda$, \AA} \\
{(1)} & {(2)} & {(3)} & {(4)} & {(5)} & {(6)} & {(7)} & {(8)}  \\  
\hline 
LS PA=88 & 2021 Nov 12 & $900\times11$ & {$1\arcsec\times6.1\arcmin$} &  0.89 & 1.4 & 3650--7250 & 5.3  \\
FPI & 2021 Dec 12 & 8580 & {$6.1\arcmin\times6.1\arcmin$} & 0.71 & 1.8 & around H$\alpha$ &  0.4  \\ 
\hline
\end{tabular}
\end{table*}

We performed the observations of the \Ha emission line with the high-resolution ($R\sim 16000$) scanning Fabry--Perot interferometer (FPI) mounted inside the SCORPIO-2 multimode focal reducer \citep{SCORPIO2}. The parameters of the FPI observations are provided in Table~\ref{tab:Obs}. Observations were carried out in the prime focus of the 6-m Russian telescope (BTA). The spectral resolution of FPI used is about $0.4$~\AA\, (corresponding to a velocity resolution of $\sim18$\kms\,  for the H$\alpha$ line) in a free spectral range (interfinge)  of $8.8$\,\AA. The operating spectral range around the \Ha emission line was cut by a narrow bandpass filter eliminating the contamination of  \NII lines around H$\alpha$ from the neighboring interference orders. 
During the observations, we consecutively obtained 40 interferograms at different gaps between the FPI plates in two fields with different position angles. The exposure for each individual channel was 240~s (for one orientation) and 120-300~s (for another orientation). The longer 300~s exposures were taken for the central channels where emission from the target was detected during the observations in the first orientation.  

The data reduction was performed using the software package running in the \textsc{idl} environment \citep{Moiseev2015, Moiseev2021}. It included primary reduction, air-glow lines subtraction, photometric and seeing corrections, and wavelength calibration. The data cubes for both field orientation were combined then into a single data cube. 

The \Ha line profiles were analyzed by fitting a single-component Voigt profile, which effectively models the emission line convolved with the instrumental profile of the FPI \citep[see][]{Moiseev2008}. This fitting procedure provides estimates of the flux, line-of-sight velocity, and intrinsic velocity dispersion (corrected for instrumental broadening) for each spatial element of the data cube. Regions with low signal-to-noise ratios ($S/N < 3$) were masked in the final maps. The typical uncertainty in velocity measurements is approximately $\sim$9\kms~ at this $S/N$ threshold, decreasing to about $\sim$2\kms~ for $S/N = 10$ \citep{Moiseev2015}.

\subsection{Long-slit spectroscopic observations}
\label{sec:longslit}

The long-slit spectra  were obtained with the same SCORPIO-2 multi-mode instrument at the BTA telescope. 
We used the  grism VPHG1200@540, which covers the wavelength range 3650-7250~\AA\, with a typical spectral resolution of 5.3~\AA\ (as estimated from the FWHM of air-glow emission lines). The slit width was 1~arcsec. The exposure time was in total 9900 s. The parameters of spectroscopic observations are also provided in Table~\ref{tab:Obs}, \rev{the position of the slit is shown on Fig.~\ref{fig:line_ratios}}.

The data were processed using a standard reduction pipeline written in Python, specifically designed for SCORPIO-2 long-slit observations. Key steps in the reduction process included bias subtraction, line curvature and flat-field corrections, wavelength calibration, and air-glow line subtraction. Each individual exposure was reduced independently and then combined with cosmic-ray rejection applied. Wavelength calibration was performed using a He-Ne-Ar lamp spectrum obtained during the observations. Flat-field correction utilized a set of LED lamps \citep[for details on the calibration system, see][]{Afanasiev2017}. Absolute flux calibration was achieved using the spectrophotometric standard stars.

The emission-line spectra were obtained by subtracting the best-fit stellar population models from the observed spectra. To subtract the spectrum of an underlying stellar population, we performed modeling using the \textsc{ULySS} package \citep{Koleva2009} with Vazdekis-Miles models \citep{Vazdekis1999}. Emission lines were then fitted with Gaussian profiles to determine their fluxes and widths. This fitting was performed using our custom software package developed in the \textsc{IDL} environment, utilizing the \textsc{mpfit} non-linear least-squares fitting routine \citep{mpfit}. All measured flux ratios were corrected for reddening based on the derived Balmer decrement, using the reddening curve from \cite{Cardelli1989} parameterized by \cite{Fitzpatrick1999}. The final uncertainties in the measured line fluxes were estimated by quadratically combining the errors propagated through all reduction steps with the fitting uncertainties returned by \textsc{mpfit}.

\section{Results}
\label{sec:results}

\subsection{Morphology and kinematics of the gaseous disk}
\label{sec:HI}

\begin{figure*}
	\includegraphics[width=2\columnwidth]{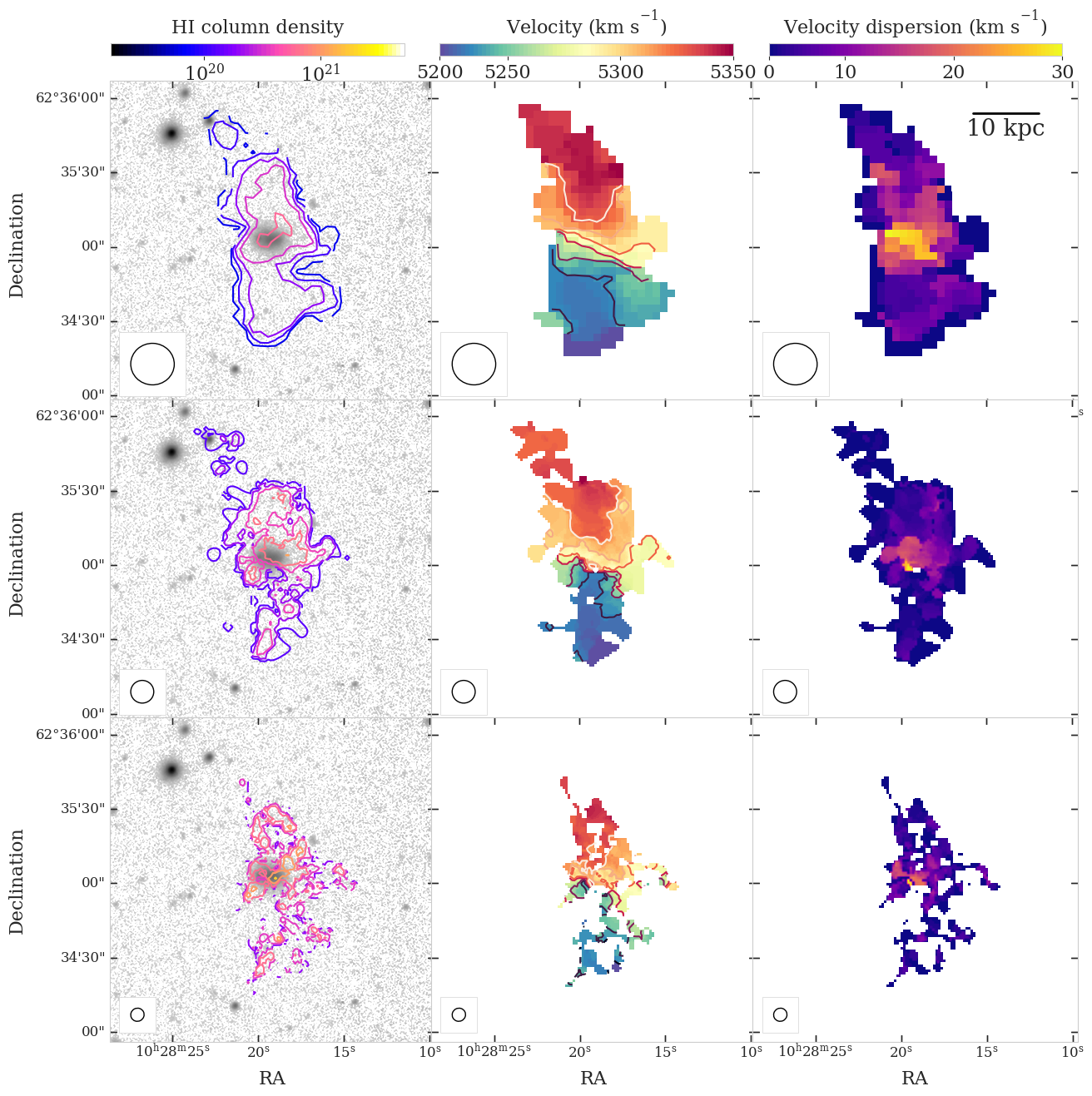}
  \caption{The \HI\ moment maps for VGS~12 at three distinct resolutions: $\sim$~17$\arcsec$ (top row), $\sim$~9$\arcsec$ (middle row), and $\sim$~5$\arcsec$ (bottom row). Left panels show the \HI\ contours overlaid on DECaLS g-band optical images. 
  The plotted \HI\ contours, starting at 2$\sigma$ level, are 3.7$\times$10$^{19}$  $\times$  ($\sqrt{3})^{n}$~\atoms\ , n=0, 1, 2, 3,... at 17$\arcsec$, 7.4$\times$10$^{19}$  $\times$  ($\sqrt{3})^{n}$~\atoms\ , n=0, 1, 2, 3,... at 9$\arcsec$, and 20.3$\times$10$^{19}$  $\times$  ($\sqrt{3})^{n}$~\atoms\ , n=0, 1, 2, 3,... at 5$\arcsec$. Middle panels display velocity fields (isovelocity contours are spaced at intervals of 20 \kms), while right panels present  Moment 2 maps. The Gaussian beam is shown on the bottom left of each map.
  }
  \label{fig:Moment_maps}
\end{figure*}

We present H~\textsc{i} moment maps for VGS~12 at three resolutions ($\sim$~17$\arcsec$ in the top row, $\sim$~9$\arcsec$ in the middle row, and $\sim$~5$\arcsec$ in the bottom row) on Figure~\ref{fig:Moment_maps}. The H~\textsc{i} disk is almost perpendicular to the stellar body of VGS~12 and asymmetrical, with a warped ``tail'' at the northern periphery. The medium and high-resolution maps reveal the clumpy disk. The velocity field is smooth at all resolutions. The elevated velocity dispersion is associated with beam smearing, as we discuss below.

For the kinematical analysis of the H~\textsc{i} data we used 3D-Barolo, which is designed for fitting 3D tilted-ring models to emission-line datacubes \citep{3dBarolo}. We ran 3D-Barolo two times on the data cube with the lowest resolution $\sim$~17$\arcsec$, as it provides the best filling of the field of view. For the first 3D-Barolo run, we let it automatically estimate the initial parameters of the fit. Even with the first initial guess, we obtained reliable results. However, we also checked the results of the 3D-Barolo fit with manually adjusted initial parameters (coordinates of the center, systemic velocity, and inclination), as discussed below. In the next 3D-Barolo run, we fix the coordinates of the center and the systemic velocity to $V_{sys}=5295$\kms, which were found in the previous step. 
This step reduces the number of free parameters and provides more robust constraints on the parameters of the individual rings. The galactic inclination in the final fit changes from $i=57\deg$ (in the centermost ring) to $i=64.5\deg$ (for most rings, the inclination is around $64\deg$), and the kinematic position angle changes from $PA = -18\deg$ to $PA = 20\deg$. The position-velocity (PV) diagrams obtained with 3D-Barolo are presented in Figure~\ref{fig:PV}. 
It clearly demonstrates that a circular rotation with a flat rotation curve dominates the \HI{} gas motions. Some deviations from this rotation pattern are considered below.
Changing the inclination (in particular, fixing $i\approx40\deg$, obtained by \cite{Stanonik2009}) or/and the coordinates of the center does not change the results significantly.

We provide the results of the final fit obtained with 3D-Barolo on Figure~\ref{fig:3DBarolo_fit}, which include the observational data, the model for each moment map, and the residuals after subtracting the model from the observational data. The high velocity dispersion in the central region of the galaxy can be fully explained by the beam smearing effect and is taken into account in the 3D-Barolo model.

FPI observations of the H$\alpha$ line provide us with a zoomed-in view on the ionized gas kinematics toward the central galaxy, and the distribution of the star formation across its disk. 
The observed \Ha velocity field was analyzed using the classical tilted-ring method \citep{Begeman1989A}, as adapted by \citet{Moiseev2014} for studying ionized-gas kinematics in dwarf galaxies. The rotation center -- identified as the point of symmetry in the velocity field -- was found to coincide closely (within 1 pixel) with the center of the continuum isophotes. This position was then fixed, and the velocity field was divided into a series of narrow elliptical rings based on the initial position angle ($PA_0$) derived from the results obtained with 3D-Barolo for the HI data. The inclination was fixed for all rings to the value obtained with 3D-Barolo ($i$ = 64~deg). Within each ring, the parameters of the circular rotation model were determined by $\chi^2$ minimization: the kinematic position angle ($PA_{kin}$), rotation velocity ($V_{rot}$), and systemic velocity ($V_{sys}$). We tested for and found no significant radial variation in $V_{sys}$. The change in the position of the rotation center, or fixing $V_{sys}$ or $PA_{kin}$ do not change the results of the fitting significantly.

\begin{figure}
	\includegraphics[width=\columnwidth]{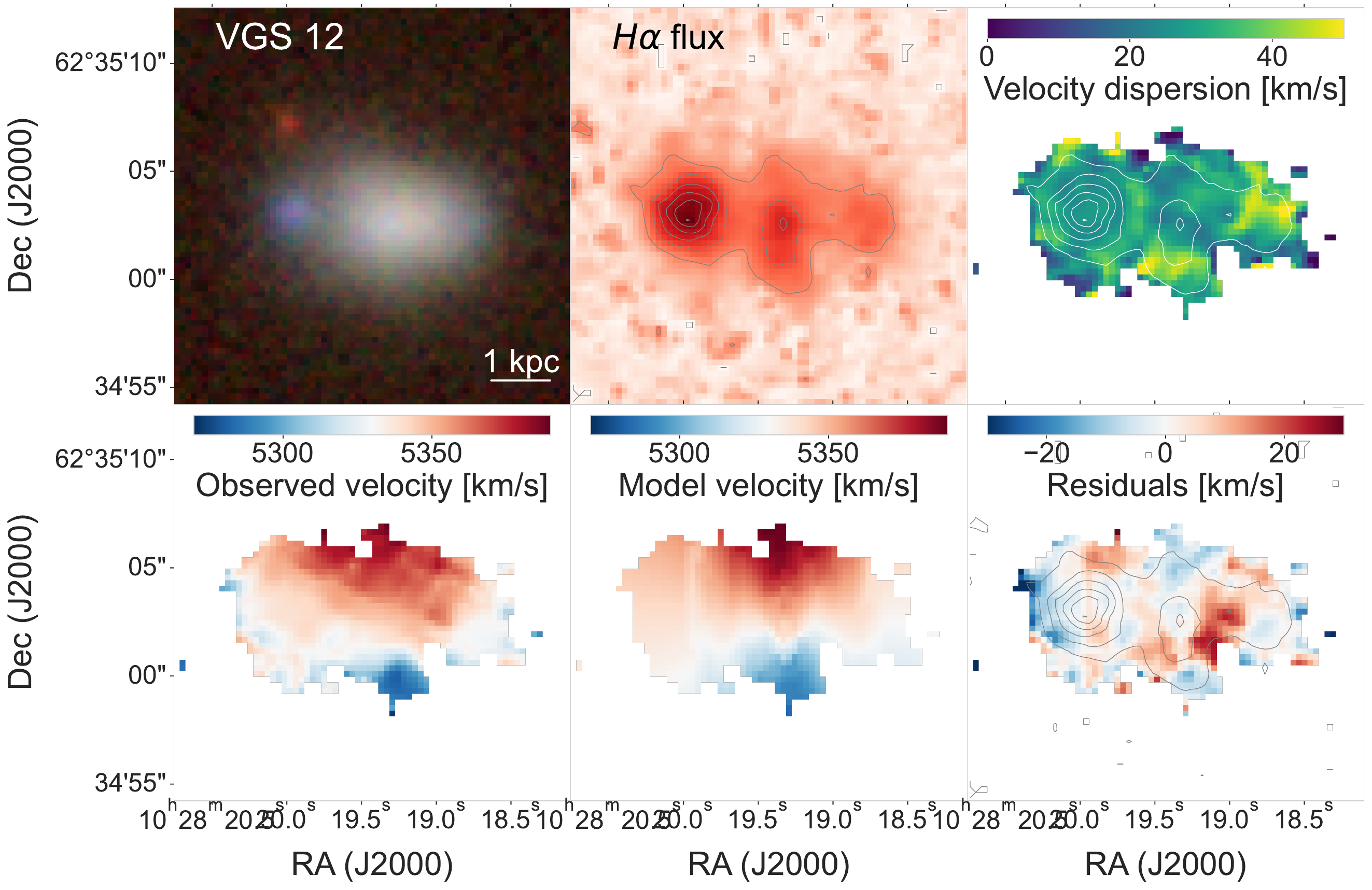}
    \caption{\textbf{Top row:} the colored Legacy Survey grz-image (left-hand panel); the map of $H\alpha$ fluxes (middle panel); the map of the ionized-gas velocity dispersions (right-hand panel). 
    \textbf{Bottom row:} the observed ionized-gas velocity field (left-hand panel); the velocity field obtained with a tilted-ring model (middle panel); the map of residuals after subtracting the model velocity field from the observed
one (right-hand panel). The \Ha contours are overlaid on the top-middle, top-right and bottom-right panels. 
    }
    \label{fig:FPI}
\end{figure}

 In Fig.~\ref{fig:FPI}, we demonstrate the multi-band image of VGS~12 from the DESI Legacy Imaging Survey \citep{Dey2019}, H$\alpha$ morphology, intrinsic velocity dispersion, line-of-sight velocity, the velocity field obtained using the tilted-ring model, and the map of residuals after subtracting the model velocity field from the observed one. No \Ha emission is detected outside the stellar disk of the galaxy. The H$\alpha$ morphology reveals a bright star-forming region coinciding with the blue clump in the optical image, as well as two other, fainter, star-forming regions. 
The ionized gas kinematics reveal a prolate rotation of the disk, i.e., the rotation along the minor axis of the galaxy.

\subsection{Gas ionization state and chemical abundances}
\label{sec:emission}

Previously, there was no information on the chemical abundances (and, in particular, oxygen abundances) for VGS~12 available in the literature, and the spectrum available in the SDSS archive for this object is too noisy to provide reliable estimates. Meanwhile, this information is crucial for verifying the metal-poor gas accretion scenario. This motivated us to perform the long-slit spectroscopic observations (see Section~\ref{sec:longslit}) and a careful analysis of the gas-phase chemical abundances. 

In Fig.~\ref{fig:line_ratios} we show the slit position overlaid on the composite Legacy Survey $grz$-image and \Ha velocity dispersion map, together with the distribution of the several key diagnostic line flux ratios along the slit. The central star-forming region reveals high \SIIHa~$\sim0.4$ flux ratio that is typically observed in diffuse ionized gas and also can be indicative of the presence of shocks (such a threshold is commonly used for identifying supernova remnants, see e.g., \citealt{Mathewson1973, Dodorico1980, Li2024}). Meanwhile, both the brightest eastern and the faintest western star-forming regions exhibit line ratios consistent with excitation due to photoionization. The \OIIIHb\, is growing and [O~\textsc{ii}]/H$\beta$ is decreasing toward the centers of these regions, which is consistent with the presence of young massive stars. However, the western region does not reveal a decrease of the ionization state at its outer edge, which could be in the case if it is ionization bounded. Together with the elevated velocity dispersion, this can be indicative of the weak outflow from this region, which in turn can be responsible for the asymmetry of the H$\alpha$ velocity field in Fig.~\ref{fig:FPI}.

\begin{figure*}
\includegraphics[width=2\columnwidth]{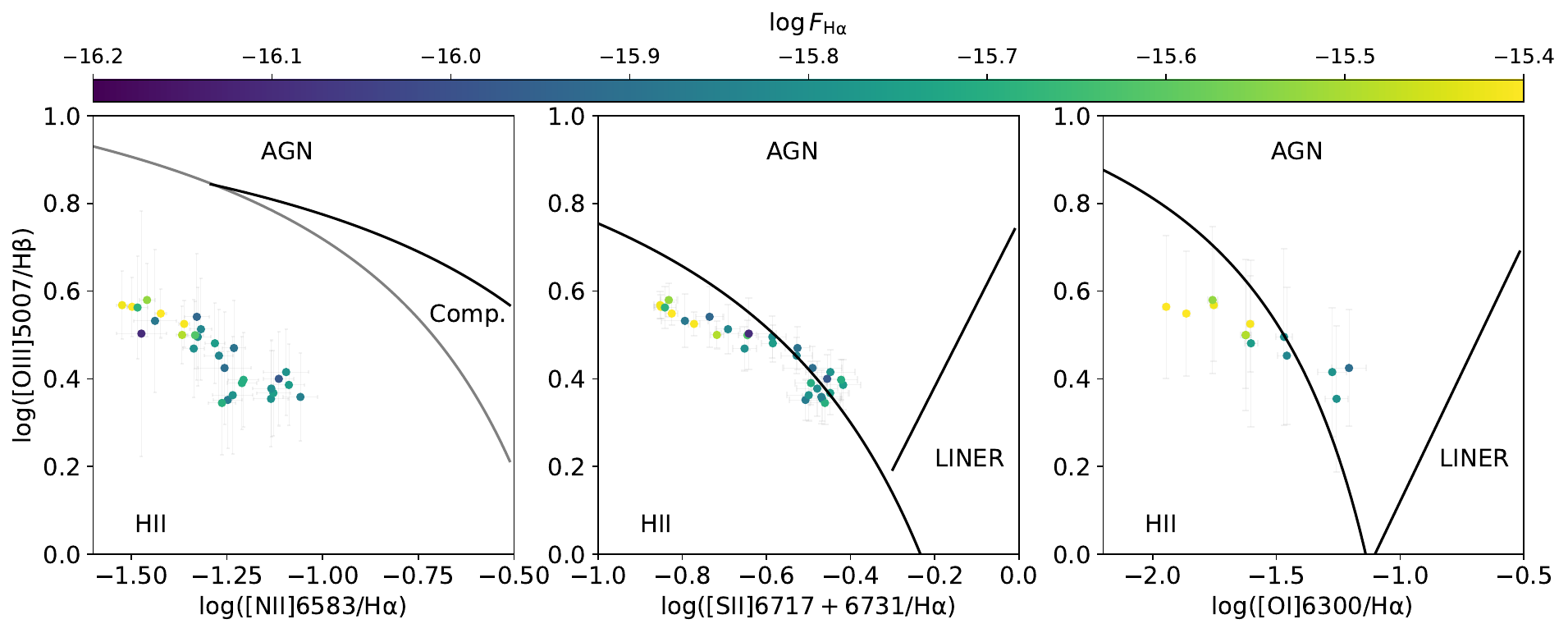}
   \caption{Diagnostic BPT diagrams. Black and gray curved lines separating the areas of different mechanisms of excitation are from \cite{Kewley2001} and \cite{Kauffmann2003}, respectively, while the straight line is from \cite{Kewley2006}. The colors of the symbols correspond to \Ha brightness.}
   \label{fig:BPT}
\end{figure*}

\begin{figure}
	\includegraphics[width=\columnwidth]{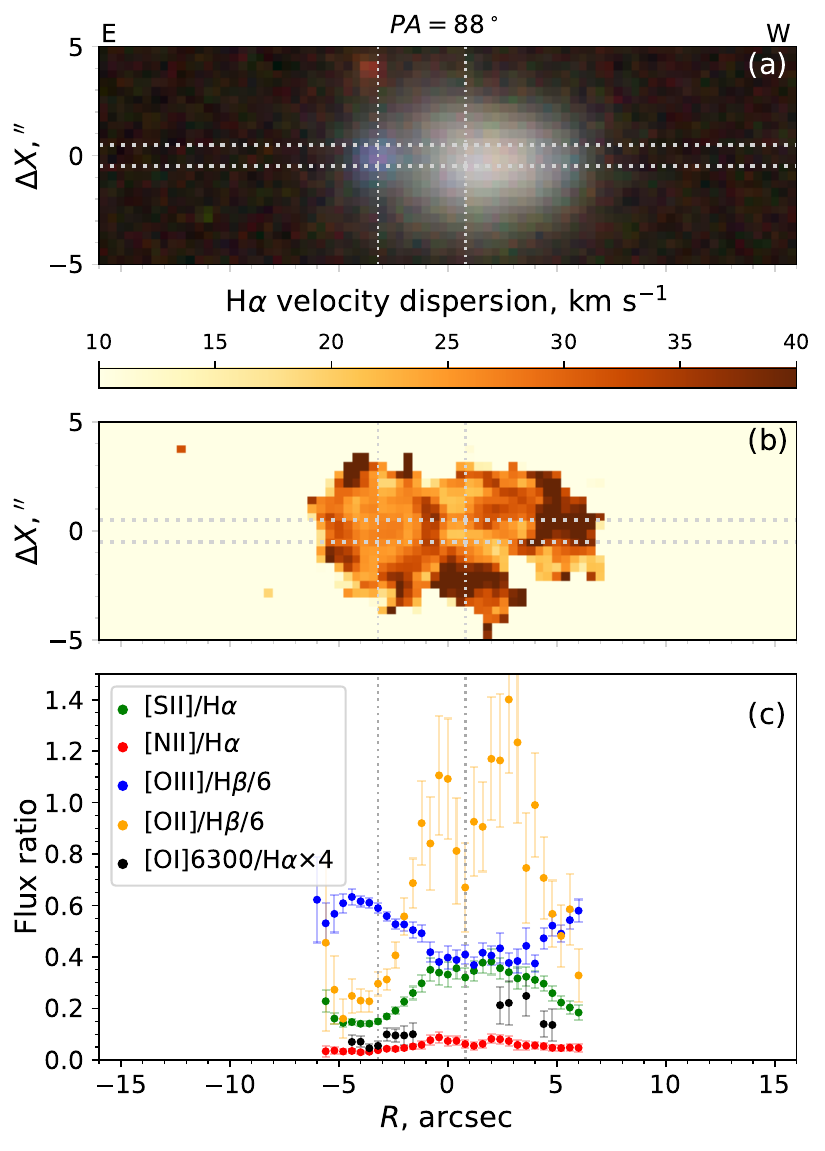}
    \caption{The colored Legacy Survey grz-image (the top panel), map of \Ha velocity dispersion (the middle panel), distribution of the emission-line fluxes ratios (the bottom panel) along the slit PA = 88. The horizontal dashed lines in the top and middle panels show the position of the slit overlaid. The vertical dotted lines denote the positions of the galaxy center and the bright SF clump on the disk periphery.}
    \label{fig:line_ratios}
\end{figure}

We also show the locations of the regions intersected by the slit on the classical BPT diagnostic diagram \citep{BPT} on Figure~\ref{fig:BPT}. The demarcation known as the ``maximum starburst line'' \citep{Kewley2001} separates regions where the emission is consistent with photoionization by young, massive stars -- indicative of ongoing star formation -- from those where additional excitation mechanisms (e.g., shocks) are likely dominant. Data points that fall between the \citet{Kewley2001} and \citet{Kauffmann2003} curves (shown in gray in Fig.~\ref{fig:BPT}) are indicative of composite ionization sources. The regions are color-coded by the \Ha flux, and the brightest regions, which are the eastern and western star-forming regions and the galactic center, are consistent with photoionization by massive stars. On the other hand, the faint \Ha regions (those between the \rev{eastern} and western SF regions and the center) show signatures of the excitation mechanisms other than photoionization by massive stars. These regions also show elevated \OIIHb~ values and normal gas velocity dispersion (see Fig.~\ref{fig:line_ratios}), which would be consistent with a high contribution of diffuse ionized gas (DIG) and unlikely to be significantly affected by shocks.

\begin{figure*}
	\includegraphics[width=2\columnwidth]{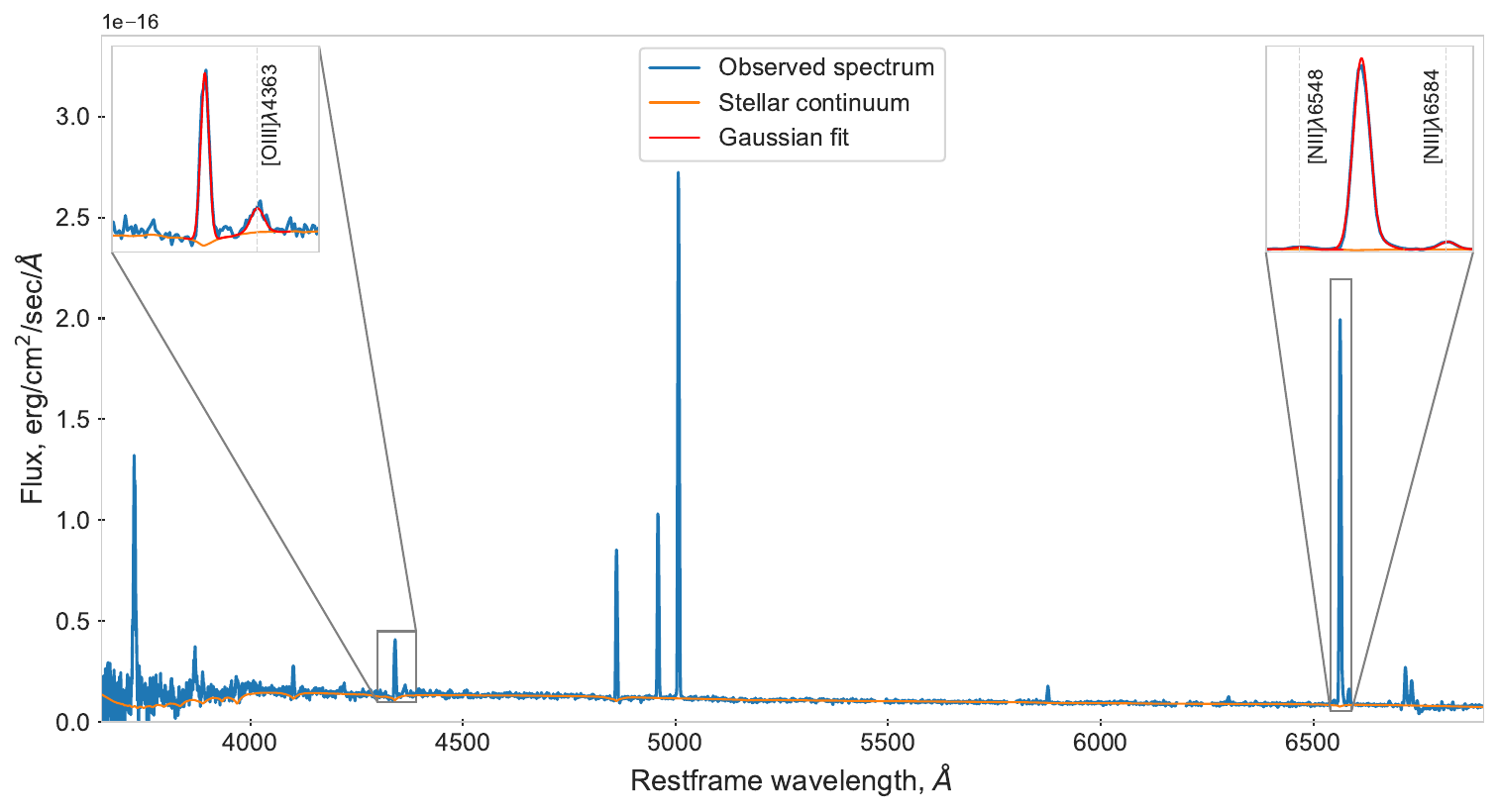}
    \caption{The spectrum of the brightest star-forming region in VGS~12. The observed spectrum at restframe wavelength is shown in blue, the continuum is in orange, the Gaussian fits for H$\gamma$ + \OIII$\lambda$4363 and \Ha + \NII$\lambda\lambda$6548,6584 lines are shown in zoom-in panels in red.}
    \label{fig:spectra}
\end{figure*}

We were able to measure the temperature-sensitive \OIII~4363\AA~line in the brightest star-forming knot on the eastern periphery of the optical disk. It allows us to measure the electron temperature $T_e$ based on the \OIII~4363\AA\ and 
5007\AA\ line ratio (see e.g., \citealt{Izotov2006}) and exploit the so-called $T_e$ (or ``direct'') method to determine the abundance of different ions. We show the spectra for this brightest region in Figure~\ref{fig:spectra}.

The ratio of [S~\textsc{ii}] 6717~\AA\ and 6731~\AA\ lines that is generally used for electron density estimates, corresponds to the low density regime ($n_e < 100\, cm^{-3}$). As was discussed in, e.g., \cite{MendezDelgado2023}, in this regime the $T_e$ diagnostic is not sensitive to the density. After calculating the electron density and temperature we derived the oxygen abundance $\mathrm{12+\log(O/H)}$ and the relative abundances of N/O, and Ne/O with the \textsc{PyNeb} package \citep{Luridiana2015} 
relying on the measurements of [O~\textsc{ii}]~3727,3729~\AA, [O~\textsc{iii}]~5007~
\AA, [N~\textsc{ii}]~6584~\AA, and [Ne~\textsc{iii}]~
3869\AA\ emission lines. Ionization correction factors from \citet{Izotov2006} were applied to the measured ionic abundances of N, S, and Ne to account for contributions from unobserved ionization states. The line fluxes, physical parameters and final abundances are given in Table~\ref{t:Intens_BTA3}.

\begin{table}
\centering{
\caption{
Measured and corrected line intensities, and derived oxygen abundances \rev{in the eastern region (RA=10:28:19.9, Dec=62:35:03.0)}}
\label{t:Intens_BTA3}
\vspace{0.1cm}
\begin{tabular}{l|c|c} \hline
\rule{0pt}{10pt}
$\lambda_{0}$(\AA) Ion     & F($\lambda$)/F(H$\beta$)&I($\lambda$)/I(H$\beta$)   \\ \hline
3727\ [O\ {\sc ii}]\               & 2.139$\pm$0.173 & 2.195$\pm$0.178 \\
3868\ [Ne\ {\sc iii}]\             & 0.301$\pm$0.031 &  0.308$\pm$0.032 \\
3970\ H$\epsilon$\                  & 0.193$\pm$0.030 & 0.197$\pm$0.031  \\
4101\ H$\delta$\                    & 0.204$\pm$0.015 & 0.207$\pm$0.015\\
4340\ H$\gamma$\                    & 0.385$\pm$0.024 & 0.389$\pm$0.024  \\
4363\ [O\ {\sc iii}]\               & 0.084$\pm$0.014 & 0.085$\pm$0.014  \\
4861\ H$\beta$\                     & 1.000$\pm$0.011 & 1.000$\pm$0.011  \\
4959\ [O\ {\sc iii}]\               & 1.156$\pm$0.011 & 1.153$\pm$0.011 \\
5007\ [O\ {\sc iii}]\               & 3.468$\pm$0.018 & 3.456$\pm$0.018  \\
5876\ He\ {\sc i}\                  & 0.109$\pm$0.005 & 0.107$\pm$0.005   \\
6300\ [O\ {\sc i}]\                & 0.055$\pm$0.006 & 0.054$\pm$0.006 \\
6548\ [N\ {\sc ii}]\                & 0.038$\pm$0.004 & 0.037$\pm$0.004  \\
6563\ H$\alpha$\                    & 2.950$\pm$0.012 & 2.860$\pm$0.012   \\
6584\ [N\ {\sc ii}]\                & 0.117$\pm$0.005 & 0.113$\pm$0.005  \\
6678\ He\ {\sc i}\                 & 0.030$\pm$0.004 &  0.029$\pm$0.004  \\
6717\ [S\ {\sc ii}]\                & 0.298$\pm$0.006 & 0.288$\pm$0.006   \\
6731\ [S\ {\sc ii}]\                & 0.212$\pm$0.005 & 0.205$\pm$0.005  \\
\hline
				                   & \multicolumn {2}{c}{~}             \\ 
E(B-V)                                 & \multicolumn{2}{c}{0.03 $\pm$ 0.01}                        \\
F(H$\beta$), 10$^{-16}$ erg s$^{-1}$ cm$^{-2}$\                                     & \multicolumn{2}{c}{3.45 $\pm$ 0.04}                        \\
$T_{\rm e}$(OIII)(K)\                            & \multicolumn{2}{c}{16831 $\pm$ 1430}                      \\
$T_{\rm e}$(OII)(K)\                             & \multicolumn{2}{c}{14630 $\pm$ 442}                      \\
log(O$^{+}$/H$^{+}$)\                 & \multicolumn{2}{c}{7.29 $\pm$ 0.06}                     \\
log(O$^{++}$/H$^{+}$)\                & \multicolumn{2}{c}{7.45 $\pm$ 0.09}                      \\
log(N$^{+}$/H$^{+}$)\                 & \multicolumn{2}{c}{5.98 $\pm$ 0.04}                     \\
log(Ne$^{++}$/H$^{+}$)\                & \multicolumn{2}{c}{6.77 $\pm$ 0.11}                      \\
12+log(O/H)($T_{e}$)\                                 & \multicolumn{2}{c}{7.67 $\pm$ 0.07}                       \\
   \hline
\end{tabular}}
\end{table}

To estimate the uncertainty of the adopted method we applied it to Monte-Carlo simulated synthetic observations with the line fluxes randomly distributed around the measured ones with a standard deviation of the probability distributions equal to the measured uncertainties of the fluxes. The value of $\mathrm{12+\log(O/H)} = 7.67\pm0.07$ and $\mathrm{\log(N/O)} = -1.29\pm0.04$ obtained from the measured fluxes agrees well with the most probable value estimated with the Monte-Carlo simulations ($\mathrm{12+\log(O/H)} = 7.67\pm0.07$ and $\mathrm{\log(N/O)} = -1.30\pm0.04$, respectively), as shown on Figure~\ref{fig:MC_direct}. 

\rev{To probe the metallicity variations across the disk, we also calculated the oxygen abundances using the strong emission line calibration from \cite{Dopita2016}, and obtained the values $\mathrm{12+\log(O/H)} = 7.67\pm0.08$, $7.80\pm0.09$, and $7.71\pm0.15$ in the eastern, central, and western regions, respectively. However, since this diagnostic involves nitrogen lines, it should be used with caution in cases of possible nitrogen overabundance, which may be the case for VGS~12 (see Sec.~\ref{sec:discussion_chem_properties}). For the subsequent analysis, we use the oxygen abundance derived with $T_e$ method, as it is generally considered to be the most reliable.}

\section{Discussion}
\label{sec:discussion}

\subsection{Chemical properties of VGS~12 in the context of its evolution}
\label{sec:discussion_chem_properties}

VGS~12 was found in the course of the Void Galaxy Survey \citep{Stanonik2009, VGS_pilot_Kreckel2011, VGS_full_Kreckel2012} that aimed at studying the properties of galaxies that reside in voids. The work by \cite{Stanonik2009} was focused on the H~\textsc{i} properties of VGS~12, and the authors suggested the cold gas accretion scenario for this galaxy based mainly on the position of the galaxy in the wall between two voids and the orientation of H~\textsc{i} and optical disks with respect to each other and to the wall. The H~\textsc{i} disk appears to be perpendicular to the optical one, it has no optical counterpart, with a mass comparable to the stellar mass of the galaxy, suggesting the recent slow accretion of the gas. However, an important piece of the puzzle that has been missing until now is the chemical properties of the gas. 

\begin{figure}
	\includegraphics[width=\columnwidth]{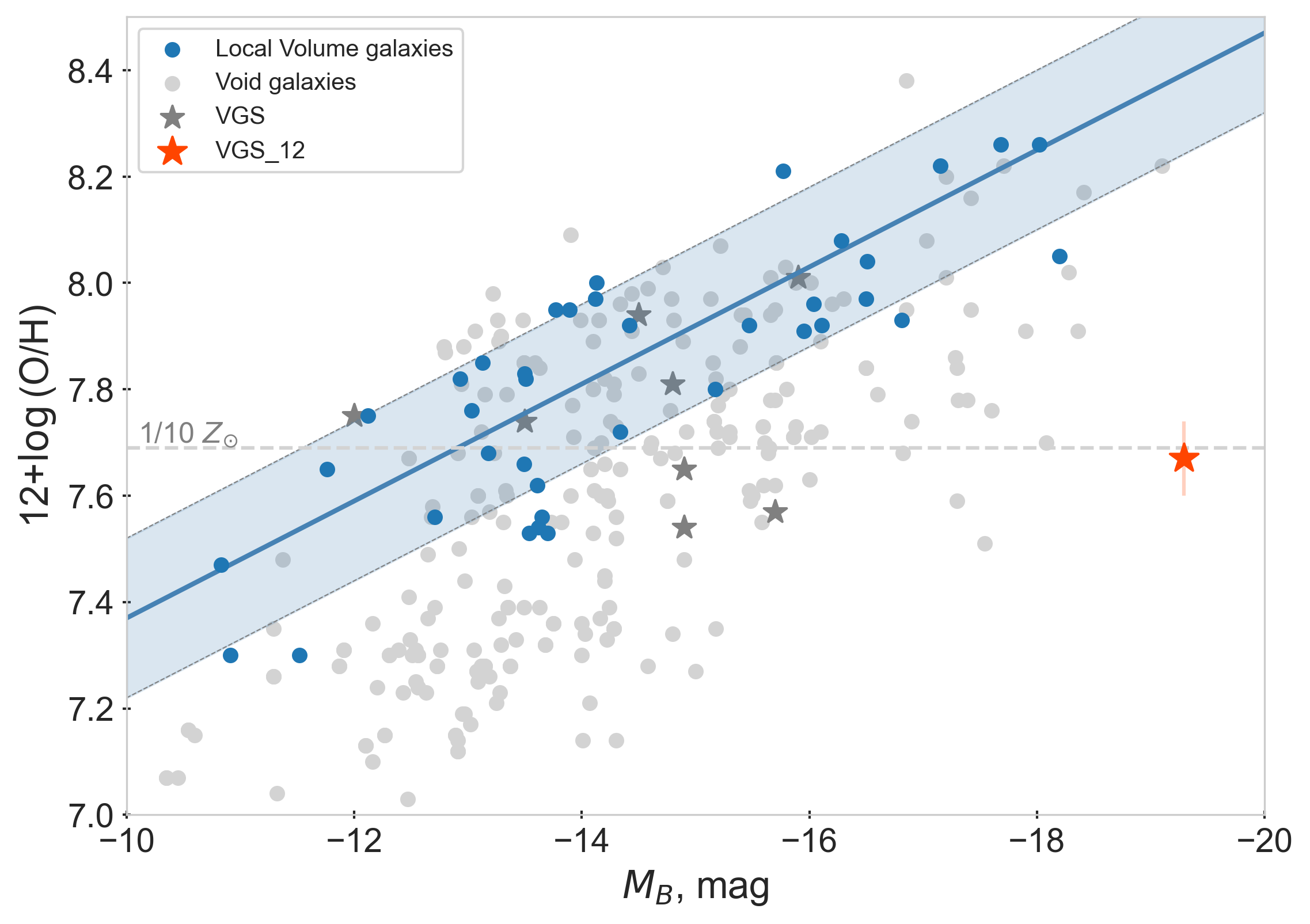}
    \caption{Relation between $\log$(O/H) and the absolute blue magnitude $M_{\rm B}$ for void galaxies (gray symbols) and Local Volume sample from \citet{Berg12} (blue symbols). The light gray circles show void galaxies from \cite{Eridanus, LC7, XMP_SALT, XMP_BTA_2021, XMP_SALT2024}. The dark gray stars denote galaxies from the Void Galaxy Survey with measured O/H from paper \cite{Kreckel2015}. VGS~12 galaxy is shown as \rev{the red} star. The solid line shows the linear regression for the control sample from the Local Volume by \citet{Berg12}. Two dashed lines on both sides of the reference line show the r.m.s. deviation of their sample from the linear regression (0.15 dex).  The horizontal dotted black line marks the value 12+$\log$(O/H)=7.69 which corresponds to Z\sunn/10 for Z\sunn\ from \citet{AllendePrieto2001} and \citet{Asplund09}.}
    \label{fig:LZ}
\end{figure}

On Figure~\ref{fig:LZ} we show the position of VGS~12 on the ``metallicity -- luminosity'' diagram and compare it with a sample of Local Volume galaxies and the corresponding relation from \cite{Berg12}. We also overplot the positions of void galaxies from Void Galaxy Survey with measured oxygen abundances \citep{Kreckel2015} and from papers by \cite{Eridanus, LC7, XMP_SALT, XMP_BTA_2021, XMP_SALT2024}.

In general, galaxies follow the ``metallicity -- luminosity'' relation \citep[e.g.,][]{Tremonti2004,Berg12,Curti2020} -- less luminous (and therefore less massive) galaxies have lower metallicity. It is generally thought that this dependence is governed by the internal processes in galaxies, such as more efficient ejection of heavy elements by galactic winds \citep[see, e.g.,][]{Tremonti2004,Finlator2008,Chisholm2018, Sanders2021}, as well as a lower efficiency of star formation and production of metals in low-mass galaxies \citep{Calura2009,Brooks2007}, the variations in initial mass function \citep{Koppen2007}, or the combination of these processes. The accretion of the metal-poor gas (or, rather, the increase of the accretion rate), including cold accretion from the filaments \citep[e.g.,][and references therein]{DeLucia2020, Ceverino2016, SA2014} and pulling the gas from the periphery toward the center during interactions or mergers of galaxies \citep[e.g.,][]{Bekki2008, Ekta10}, may lead to deviations from the ``metallicity -- luminosity'' relation to lower oxygen abundances. Local drops in oxygen abundances coinciding with the SF regions that were found in the so-called tadpole (or cometary) galaxies with off-center star-forming clumps are also usually interpreted as the effect of pristine gas infall \citep[e.g.,][]{Lagos2018, SA2013}. It's interesting to note that VGS~12 also reveals bright star forming clump on its periphery (see Figures~\ref{fig:FPI},\ref{fig:line_ratios}), although our spectroscopic data do not allow us to obtain reliable estimates of metallicity outside the brightest clump or to determine the distribution across the disk. \rev{It is worth noticing that VGS~12, being a relatively bright galaxy, can exhibit a metallicity gradient, and measuring the oxygen abundance at the periphery of the disk may also lead to some deviation from the metallicity-luminosity relation toward lower oxygen abundance. However, given the typical metallicity gradients found in previous studies \citep[e.g.,][]{Sanchez2014,Ho2015,Belfiore2017}, this effect can explain only up to $\sim$0.1~dex difference in the case of VGS~12, which is in agreement with the variations of oxygen abundance across the disk obtained with strong-line method (see Sec.~\ref{sec:emission}).}

Another important diagnostic is the N/O versus O/H diagram. Nitrogen can be produced as both the primary and secondary element, with secondary nitrogen produced mostly in intermediate-mass stars and is delayed relative to oxygen production. Primary production of nitrogen is due to the evolution of massive stars, and  dominates at low metallicities and leads to a plateau on the N/O versus O/H relationship at log(N/O) $\sim -1.5$, whereas secondary production starts to dominate at intermediate metallicities and is reflected in an increase in N/O with increasing O/H \citep[e.g.,][]{Vila-Costas1993, Henry2000,vanZee2006,Berg12}. The observed scatter of N/O is significant, as discussed in a number of previous studies \citep[e.g.,][]{Belfiore2015,Berg2020}, and can be associated with bursts of star formation \citep{Torres-Papaqui2012} and star formation histories \citep{Berg2020}, galactic fountains \citep[see the discussion in][]{Belfiore2015}, pollution by nitrogen-rich winds from population of Wolf-Rayet stars \citep[e.g.,][]{Brinchmann2008}, and the inflows of metal-poor gas \citep{Koppen2005}.  
In particular, dilution by the pristine gas mentioned above should lead to a nitrogen-to-oxygen overabundance for a given O/H or N/H, as it decreases both O/H and N/H, but does not affect the N/O ratio. 

\begin{figure}
	\includegraphics[width=\columnwidth]{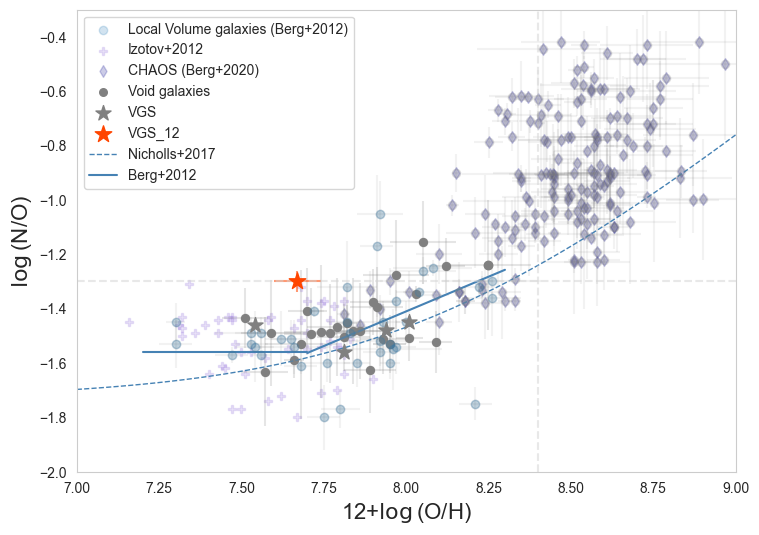}
    \caption{Relation between $\log$(N/O) and $\log$(O/H). The dark gray circles show void galaxies from \cite{Eridanus, LC7, XMP_SALT, XMP_BTA_2021, XMP_SALT2024}. The dark gray stars denote galaxies from the Void Galaxy Survey with measured O/H from paper \cite{Kreckel2015}. VGS~12 galaxy is shown as \rev{the red} star. The purple crosses denote galaxies from \cite{Izotov2012}, blue diamonds -- HII regions in NGC~628, NGC~3184, NGC~5194, NGC~5457 from CHAOS survey from \cite{Berg2020}. The blue line shows the stellar ``$\log\mathrm{(N/O)}$ -- $12+\log\mathrm{(O/H)}$'' relation from \cite{Nicholls2017}. The horizontal dashed line marks the measured $\log$(N/O) for VGS~12; the vertical dashed line marks $12+\log\mathrm{(O/H)}$ expected for the given $M_B$ estimated using the relation from \cite{Berg12}.}
    \label{fig:NO}
\end{figure}

In Fig.~\ref{fig:NO} we show $\log$(N/O) versus 12+$\log$(O/H) for Local Volume galaxies from \cite{Berg12}, NGC~628, NGC~3184, NGC~5194, NGC~5457 from CHAOS survey \citep{Berg2020}, low metallicity emission line galaxies from \cite{Izotov2012}, void galaxies from the Void Galaxy Survey with measured O/H from \cite{Kreckel2015} and from \cite{Eridanus, LC7, XMP_SALT, XMP_BTA_2021, XMP_SALT2024}. The position of the VGS~12 galaxy is shown by the red star.
Our analysis of the spectral data reveals that VGS~12 appears to be an outlier on both the ``metallicity -- luminosity'' relation and ``N/O versus O/H'' diagram. At a given $M_B = -19.29$ the oxygen abundance is expected to be $12+\log\mathrm{(O/H)} \sim 8.4$ (according to the ``metallicity -- luminosity'' relation from \cite{Berg12}), so the measured gas-phase oxygen abundance ($12+\log\mathrm{(O/H)} = 7.67\pm0.07$) is 0.7~dex lower than expected. At the same time, its relatively high N/O (compared to most of the galaxies with similar oxygen abundance) is consistent with the oxygen abundance expected for the given $M_B$ (that is, to $12+\log\mathrm{(O/H)} \sim 8.4$). The accretion of the metal-poor gas, suggested by \cite{Stanonik2009}, would naturally explain the outlying position of VGS~12 on both the ``metallicity -- luminosity'' relation and ``N/O versus O/H'' diagram, and our findings strongly support this scenario.

The simultaneous local drops of O/H and relatively high N/O abundance ratios are interpreted as a signature of possible metal-poor gas accretion in some observational studies \citep[for the review see, e.g.,][and references therein]{SA2014}. For example, \cite{Luo2021} identified high N/O abundances in regions of anomalously low metallicity in nearby star-forming galaxies in the SDSS IV Mapping Nearby Galaxies at Apache Point Observatory survey \citep[MANGA,][]{Yan2016, Bundy2015}. The authors concluded that the locations of these regions on the N/O versus O/H plane are consistent with a model where metal-poor gas is accreted from outside the galaxy. Similarly, \cite{Valle-Espinosa2023} suggests that the accretion of metal-poor gas is responsible for the starbursts in CGCG~007-025 dwarf galaxy. They found a homogeneous distribution of N/O together with a decrease of O/H at the position of the SF clumps for this galaxy. 

VGS~12 provides a unique example in which such chemical indicators of diluting by metal-poor gas accretion are accompanied also by clear kinematic signatures of  inflowing gas from the surrounding intergalactic medium.

\subsection{Properties of the gaseous disk}
\label{sec:disc_gas}

VGS~12, although not the only galaxy with a gaseous polar disk, represents a subclass of rare objects, that is poorly explored so far. Its detailed study can shed light on the formation scenario of such galaxies, and our chemical abundance analysis favors cold gas accretion from filaments.

The polar disk in VGS~12 was discovered by \cite{Stanonik2009} via its H~\textsc{i} content using the data from the Westerbork Synthesis Radio Telescope (WSRT) with a beam size of 22$\arcsec$.1 × 19$\arcsec$.5. \cite{Stanonik2009} also found a hint of a warp in the outer parts of the H~\textsc{i} disk. With VLA data, we can investigate the H~\textsc{i} morphology and kinematics of VGS~12 with higher resolution (from the lowest resolution $\sim$~17$\arcsec$ to the highest resolution $\sim$~5$\arcsec$). With this resolution, the H~\textsc{i} disk shows a clumpy structure and clear asymmetry between the southern and northern parts. The kinematical position angle changes significantly, from $PA = -18\deg$ to $PA = 20\deg$. 

The best-fit model reveals high residual velocities after subtracting the model from the observed velocity field (see Figure~\ref{fig:3DBarolo_fit}) in the northern part of the galaxy. We were not able to obtain a uniformly good fit for the velocity field for both southern and northern parts simultaneously by changing the initial parameters of the 3D-Barolo fit (coordinates of the center, systemic velocity, or inclination).  
Both the morphological asymmetry and the relatively high residual velocities in the northern part of VGS~12 are likely related to the recent accretion event and can be explained by the unsettled state of the gaseous disk.

As we already mentioned in Sec.~\ref{sec:HI}, the rotation of the ionized gas is prolate, i.e., it takes place along the minor galaxy axis. Such behavior is quite unusual for normal galaxies, although the H$\alpha$ velocity field is fully consistent with the rotation of the H~\textsc{i} disk observed on larger scales. The fainter western clump exhibits elevated velocity dispersion and line-of-sight velocity clearly deviating from the rotational pattern observed in the rest of the disk. Although the \HI data do not resolve the galactic disk, the \HI iso-velocities also show a similar bend at the western part of the galaxy. Therefore, such a peculiar velocity field is consistent between \Ha and H~\textsc{i}, and thus our zoom-in view with \Ha kinematics probably traces the impact of large-scale gas flows, although we cannot exclude the presence of the small-scale feedback-driven outflow in the western clump (see Section~\ref{sec:emission}).

As can be seen from Figure~\ref{fig:FPI}, the velocity dispersion of the ionized gas is elevated (up to $\sim$50\kms) in the western and southern periphery of the optical disk. These regions do not coincide with the position of the bright \HII region that is located on the eastern periphery of the disk, and can be related to the shock excitation due to the collision of the accreted gas with the gravitational potential of the central stellar disk \citep{Egorov2019}.  This also can explain the relatively high residuals after subtracting the model velocity field from the observed one near the western edge of the stellar disk.

The exceptional agreement between kinematics of neutral and ionized gas suggests that the central part of \HI polar disk is ionized by the stellar disk radiation. The observational properties of the ionized gas disk described in Sect.~\ref{sec:results} are consistent with ionization by massive stars in two star-forming regions, DIG in most of the stellar disk, and possible shock excitation in some peripheral regions. 

\rev{VGS~12, with its \HI disk perpendicular to the central stellar disk, is a polar ring galaxy (PRG). PRGs are systems that reveal a ring or disk (stellar and/or gaseous) perpendicular to the central galaxy. Polar rings are generally considered to be formed as the result of mergers \citep{Bekki1998,Bekki1997}, accretion from the companion galaxy \citep[e.g.,][]{Reshetnikov1997,Bornaud_Combes2003}, or cold accretion from gaseous filaments \citep{Brook2008,Maccio2006}, but different models and simulations favor different scenarios. Observational measurements of gas excitation and metallicity in PRGs favor the scenario of their formation as a result of interaction with companions \citep[e.g.,][]{Egorov2019}, with a few exceptions, such as NGC~4650A, where metal-poor gas accretion was shown to be a possible scenario \citep[e.g.,][]{Spavone2010}.}

\rev{There are also a number of known systems with gaseous polar structures besides VGS~12, such as the Local Group galaxy NGC6822 \citep{NGC6822_polar} and the dwarf elliptical galaxy FCC046 \citep{FCC046_polar} residing in the Fornax cluster. \cite{Serra2012} found three \HI polar disks examining the sample of 166 early-type galaxies from the ATLAS$^{3D}$ project \citep{Cappellari2011}. In a recent paper by \cite{Deg2023}, the authors also found two potential gaseous polar rings, NGC4632 and NGC6156, in the WALLABY pilot survey. They argue that the incidence of gaseous polar structures should be $\sim$1-3\%, which is consistent with new estimates obtained by \cite{Mosenkov2024} for low-brightness stellar polar structure. \cite{Smirnov2024} also found in Illustris-TNG50 cosmological simulation \citep{Pillepich2018} a number of galaxies with exclusively gas polar structures, without prominent stellar counterparts, which are particularly relevant for the case of VGS~12.}

\subsection{Gas accretion in voids}

As was mentioned in the introduction, theoretical studies and simulations predict that voids have complex substructures and contain smaller subvoids, walls and filaments  \citep[e.g.,][]{Dubinski1993, vandeWeygaert&Kampen1993, Sahni1994, Sheth04, Weyplaten2009,Cautun2014}. They should be filled with a tenuous network of filaments of metal-poor gas and dark matter, with galaxies sitting along these walls and filaments and in the nodes of the large-Scale structure where these intersect \citep{Bond1996,Gottlober2003,Sheth04,Weyplaten2009,Aragon2013,rieder13}, and matter generally flows from the inner parts of the void along the filaments toward the void boundaries delimited by walls \citep[e.g.,][]{vandeWeygaert&Kampen1993,Cautun2014,Cautun2016,Kugelwey2024}. 

The filaments in the Cosmic Web span several orders of magnitude in mass density and differ by factors of several in thickness \citep{Cautun2014}, starting from dense thick filaments connecting clusters of galaxies to tenuous tendrils inside voids. Filaments in voids are typically faint and thin low-density channels, loosely populated with haloes. More detailed discussion on the nature of the filaments in Cosmic Web can be found in \cite{Feldbrugge2025}. In particular, they found that small filaments should typically have formed earlier, whereas large filaments reveal the diversity in history and formation time, with the largest filaments still forming.

According to simulations, galaxies in voids can be fed by gas accretion from the filaments, and a steady inflow of matter can be kept there for extended periods of time \citep{Aragon2013,Aragon-Calvo2019}. In particular, \citet{Aragon-Calvo2019} presented results from N-body simulations of a halo located in a low-density region (see their Fig. 22), which may serve as a counterpart to VGS~12. They showed that the halo remains connected to gas filaments and that cold gas accretion continues at a relatively steady rate until the present time. \cite{Jaber2024} discuss that relatively low-mass void galaxies are younger and bluer and have lower metallicity than galaxies, and interpret this as effect of more frequent accretion from filaments.

The presence of large amounts of pristine gas and the ongoing gas accretion could explain the possible slower evolution of void galaxies compared to denser environments that was previously discussed in several observational studies. For example, the population of extremely metal-poor void dwarfs was found by \cite{LC2, Eridanus, XMP_BTA_2021, XMP_SALT, XMP_SALT2024}. These galaxies, similar to VGS~12, are outliers from the ``metallicity -- luminosity'' relation, blue \citep{LC4} and very gas-rich \citep{Kurapati2024_XMP}, and are good candidates for young galaxies. \cite{Cavity2023_nat}, as part of CAVITY survey \citep{CAVITY2024}, performed the spectral analysis of galaxies in different environments. They found that void galaxies reveal, on average, slower star formation histories compared to galaxies in denser environments. The authors discuss gas accretion as one of the possible mechanisms responsible for this.

The evidence of metal-poor cold gas accretion in voids has been discussed in several other studies. In \cite{VGS_full_Kreckel2012}, the authors revealed that about half of the Void Galaxy Survey galaxies that were detected in H~\textsc{i} show strongly disturbed gas kinematics and morphology. They interpret it as a sign of ongoing interactions and accretion. Another VGS~target, VGS~31 system, presents an example of possible substructure formation \citep{VGS2013}. It reside deep in the void and consists of three almost linearly aligned galaxies embedded in a common H~\textsc{i} envelope. The eastern-most component reveals the outlying polar disk with low-metallicity, that also was interpreted as possible sigh of cold gas accretion. According to simulations \citep{rieder13}, the VGS~31 galaxies didn't meet recently, but should have formed in the same protofilament.

Similar linearly aligned systems in voids, consistent with galaxies lying along large scale filaments, were discovered in other studies. \cite{CPE2017} found that UGC~3672 system, residing within the central 8\% volume of the Lynx-Cancer void, contains three gas-rich galaxies that are located inside a common H~\textsc{i} envelope. The faintest component is metal deficient ($12+\log\mathrm{(O/H)} \sim 7.0$), extremely gas-rich ($M_{HI}/L_B \sim 17$), with colors consistent with the recent ``onset'' of star formation. These unusual properties can be related to the inflow of pre-existing faint dwarf or gas clump along the filament toward the main pair of galaxies. Another example is the gas-rich metal-poor dwarf triplet J0723+36 that was found in the central region of the same Lynx-Cancer void \citep{CP2013}. 

In a more recent study by \cite{Kurapati2024} in the SARAO MeerKAT Galactic Plane Survey the Local Void was explored and two groups of galaxies, with 3 and 5 members, were found that show signs of filamentary substructure. This was interpreted as the possible ongoing growth of these galaxies along a filament inside the Void. \cite{Tully2019} in Cosmicflows-3 also revealed the filament with several galaxies along the chain that goes through the deep density minima in the Local Void.

\cite{Kurapati2024_XMP} recently found a galaxy J2103–0049 that resembles VGS12 with its properties. J2103-0049 resides in a void, it has a roundish form of the optical LSB outer body but reveals a very elongated H~\textsc{i} disk. This extremely metal-poor galaxy with $12+\log\mathrm{(O/H)} \sim 7.2$ \citep{XMP_BTA_2021} falls down from the ``metallicity -- luminosity''  relation from \cite{Berg12} by $\sim$0.6 dex. These properties also may indicate the recent accretion of the metal-poor gas from the Cosmic Web, and the authors discuss it as a possible scenario for this galaxy.

From the discussion above, it can be seen that the large-scale filamentary substructures and the gas accretion in voids that were predicted in simulations already have strong observational evidence. VGS~12 is not the only, but possibly the most striking example of this. 

\section{Conclusions}
\label{sec:conclusions}

We present the analysis for the unusual polar-disk galaxy VGS~12 utilizing the previously unpublished VLA H~\textsc{i} data, together with long-slit spectroscopic and Fabry-Perot interferometer data obtained at the Russian 6-m telescope BTA SAO RAS. For the first time, we constrain the chemical abundance in VGS~12. 

We measured gas-phase oxygen abundance $12+\log\mathrm{(O/H)}=7.67\pm0.07$  with the $T_e$ (`direct') method, and nitrogen abundance $\log\mathrm{(N/O)}=-1.29\pm0.04$. The Oxygen abundance appears to be significantly (by 0.7~dex) lower than expected for the given luminosity compared to the ``metallicity -- luminosity'' relation from \cite{Berg12}. At the same time, VGS~12 is an outlier on the N/O versus O/H diagram and reveals higher N/O than typical galaxies with similar metallicities. However, the observed N/O is in agreement with $12+\log\mathrm{(O/H)}\sim8.4$, expected for the given VGS~12 luminosity. This finding strongly supports the scenario of the metal-poor gas accretion from the adjacent void suggested by \cite{Stanonik2009}, as the dilution of ISM by the pristine gas should lower the oxygen abundance, but does not affect N/O.

The H~\textsc{i} polar disk is mapped with VLA at different resolutions ($\sim17\arcsec$, $\sim9\arcsec$, and $\sim5\arcsec$) and appears to be clumpy, showing clear morphological and kinematical asymmetry between the southern and northern parts. Such asymmetry and clumpiness can be explained by the unsettled state of the gaseous disk in the case of the recent accretion event.

The kinematics of the ionized gas in the \Ha line reveals  prolate rotation, but closely follows the kinematics of the \HI disk. It's likely that the accreted H~\textsc{i} gas is ionized by stars in the central part of the galaxy. We have not found any \Ha emission outside the optical disk of VGS~12. Neither did we find any tidal features associated with the galaxy on the optical images, which would be expected in the case of recent mergers. 

The scenario of metal-poor gas accretion from the filament of the adjacent void was suggested by \cite{Stanonik2009} based on the position and orientation of VGS~12 in the wall between two voids and the mutual orientation of \HI and optical disks. Our observational results on VGS~12 are in agreement with this scenario and strongly support it. Due to the unique circumstances of galaxy evolution in voids, we believe these represent a proper environment to search for definitive signs of pristine gas accretion.  VGS~12 provides some of the strongest observational evidence that this process can continue contributing to galaxy grown until the present day.

\begin{acknowledgements}
The authors thank Simon Pustilnik for fruitful discussion and providing useful comments. 
EE and KK gratefully acknowledge funding from the Deutsche Forschungsgemeinschaft (DFG, German Research Foundation) in the form of an Emmy Noether Research Group (grant number KR4598/2-1, PI Kreckel) and the European Research Council’s starting grant ERC StG-101077573 (“ISM-METALS"). OE acknowledges funding from the Deutsche Forschungsgemeinschaft (DFG, German Research Foundation) -- project-ID 541068876. The work of AM on the interpretation of gas kinematics and ionization (Sec.~\ref{sec:results} and Sec.~\ref{sec:disc_gas}) has been supported by Russian Science Foundation grant No. 22-12-00080. MAAC acknowledges support from CONAHCyT Ciencia de Frontera grant CF-2023-1-1971 and UNAM PAPIIT grant IN115224. \rev{RvdW acknowledges funding from EU Horizon Europe (EXCOSM, grant nr. 101159513).}

The National Radio Astronomy Observatory is a facility of the National Science Foundation operated under cooperative agreement by Associated Universities, Inc.
Observations with the SAO RAS telescopes are supported by the Ministry of Science and Higher Education of the Russian Federation. The renovation of telescope equipment is currently provided within the national project ``Science and Universities''.
\end{acknowledgements}
  \bibliographystyle{aa} 
  \bibliography{biblio} 

\begin{appendix} 
\section{The model for low-resolution ($\sim$~17$\arcsec$) data-cube obtained with $^{3{D}}$BAROLO}
$^{3{D}}$BAROLO code \citep{3dBarolo} is developed for emission-line observations and fits 3D tilted-ring models to spectroscopic data cubes. It can recover the rotation curve, estimate the intrinsic velocity dispersion even in low-resolution data and allows one to overcome the beam-smearing problem. We use the code to model our lowest-resolution ($\sim$~17$\arcsec$) HI data cube obtained with the VLA (see Sec.\ref{sec:data_VLA}, \ref{sec:HI}).
\begin{figure}[h]
	\includegraphics[width=\columnwidth]{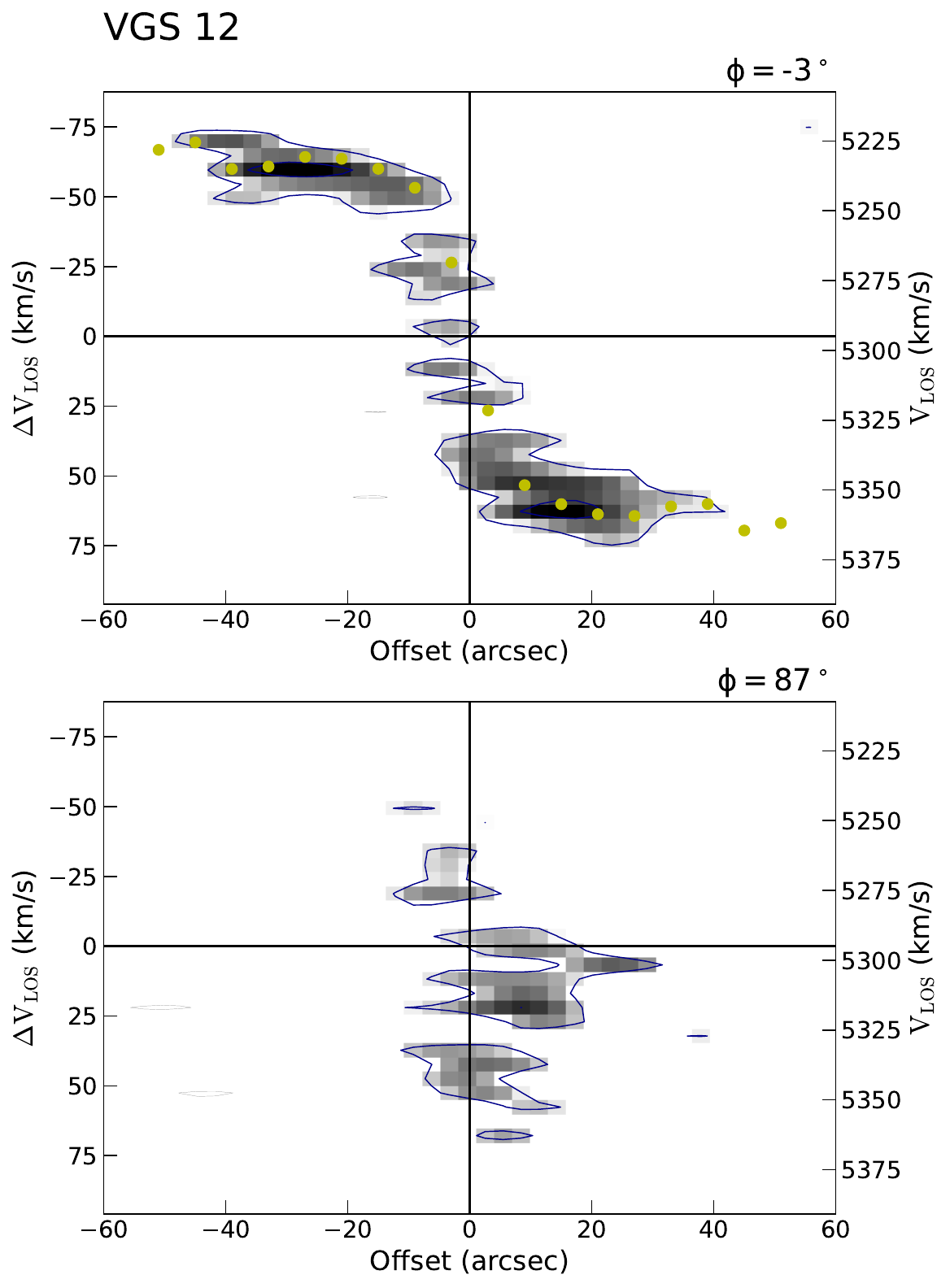}
  \caption{PV diagrams along the major (top panel) and minor (bottom plot) axes of HI disk obtained with 3D-Barolo tool \citep{3dBarolo} for the low-resolution ($\sim$~17$\arcsec$) HI data-cube. Data are represented in gray, rotation curve as yellow circles.
  }
  \label{fig:PV}
\end{figure}

\begin{figure*}[h]
\centering
\includegraphics[width=1.8\columnwidth]{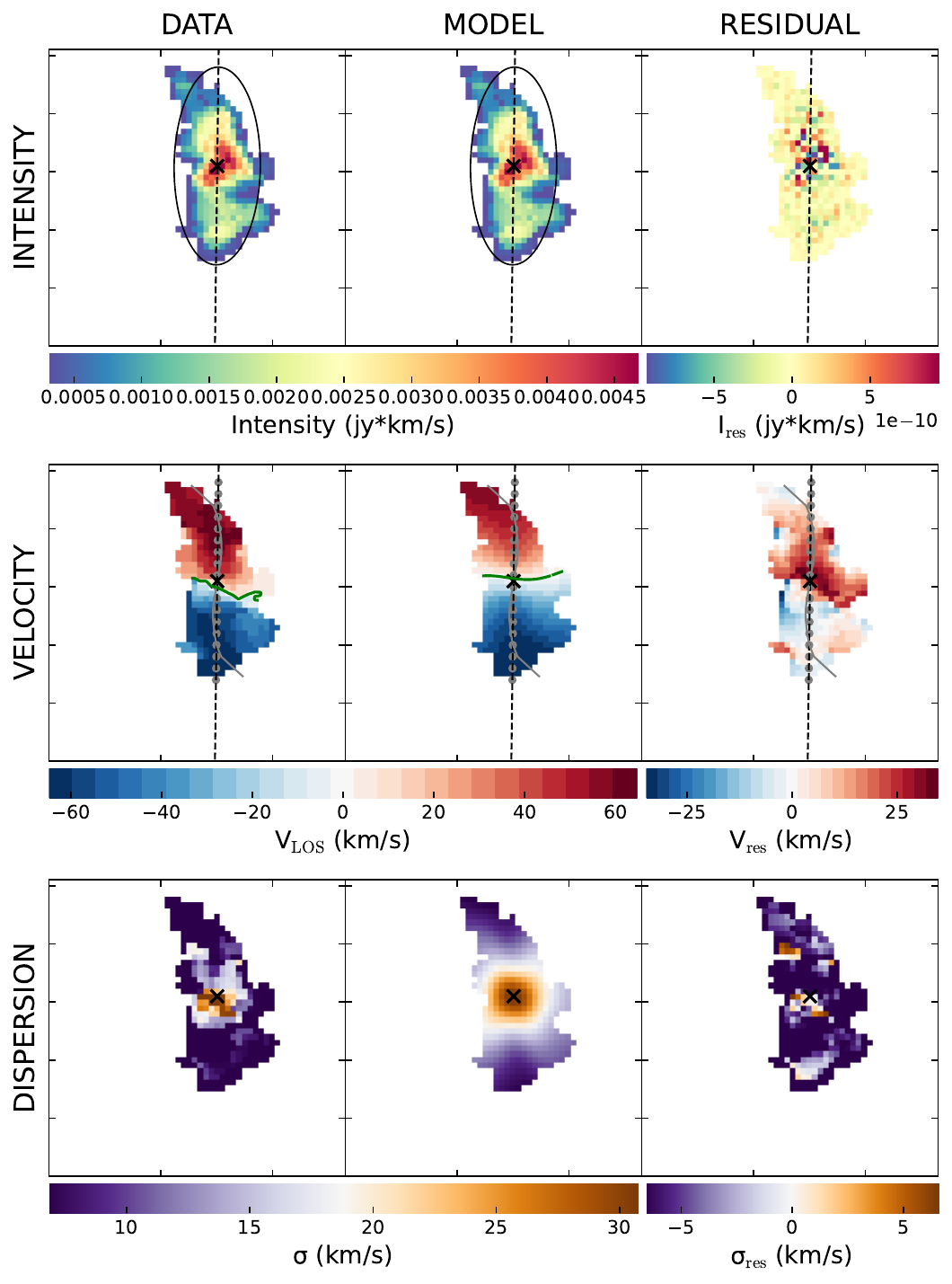}
  \caption{The model obtained with $^{3{D}}$BAROLO code \citep{3dBarolo} for low-resolution ($\sim$~17$\arcsec$) data-cube. \textbf{From left to right:} observational data, model obtained with $^{3{D}}$BAROLO, residuals after subtracting the model from the observational data. \textbf{From top to bottom:} HI line intensity, velocity field, and velocity dispersion.
  }
  \label{fig:3DBarolo_fit}
\end{figure*}

\section{The results of Monte-Carlo simulations for 12 + log(O/H) and log(N/O).}

\begin{figure*}
\includegraphics[width=1.9\columnwidth]{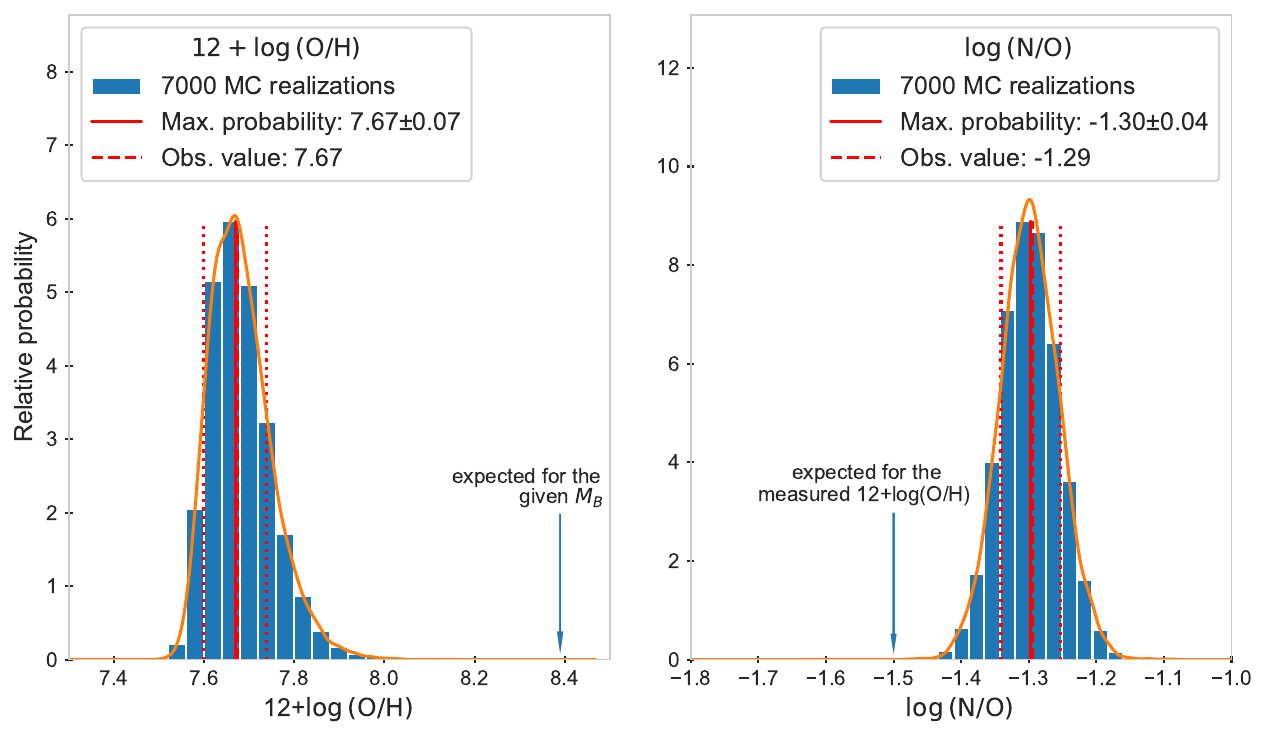}
   \caption{The results of Monte-Carlo simulations to estimate the uncertainty in our measurements of $\mathrm{12+\log(O/H)}$ (\textbf{left panel}) and $\mathrm{\log(N/O)}$ (\textbf{right panel}). The histograms show the distribution of 7000 Monte-Carlo realisations results, the red dashed line marks the value obtained with \textsc{PyNeb}, the red solid line and dotted lines mark the position of maximum probability of the Monte-Carlo results and its standard deviation, respectively. The blue arrow marks $\mathrm{12+\log(O/H)}$ expected for the given luminosity according to the ``metallicity -- luminosity'' relation be \cite{Berg12} (see Figure~\ref{fig:LZ}), and $\mathrm{\log(N/O)}$ expected for the observed oxygen abundance according to the N/O versus O/H diagram (see Figure~\ref{fig:NO}).}
   \label{fig:MC_direct}
\end{figure*}

To estimate the uncertainty in our calculation of the chemical abundances we applied our analysis  to Monte-Carlo simulated synthetic observations, where the line fluxes are randomly distributed around the measured fluxes assuming a standard deviation of the probability distributions equal to the measured uncertainties of the fluxes (Figure \ref{fig:MC_direct}).

\end{appendix}

\end{document}